\documentclass[%
reprint,
superscriptaddress,
amsmath,amssymb,
aps,
]{revtex4-2}

\usepackage{graphicx}
\usepackage{dcolumn}
\usepackage{bm}
\usepackage{hyperref}

\usepackage{txfonts}
\usepackage{braket}
\usepackage{comment}
\usepackage{color}

\begin{document}

\newcommand\red[1]{\textcolor{red}{#1}}
\newcommand\blue[1]{\textcolor{blue}{#1}}

\title{Surface wetting by kinetic control of liquid-liquid phase separation}

\author{Kyosuke Adachi}
\affiliation{Nonequilibrium Physics of Living Matter RIKEN Hakubi Research Team, RIKEN Center for Biosystems Dynamics Research, 2-2-3 Minatojima-minamimachi, Chuo-ku, Kobe 650-0047, Japan}
\affiliation{RIKEN Interdisciplinary Theoretical and Mathematical Sciences Program, 2-1 Hirosawa, Wako 351-0198, Japan}

\author{Kyogo Kawaguchi}
\affiliation{Nonequilibrium Physics of Living Matter RIKEN Hakubi Research Team, RIKEN Center for Biosystems Dynamics Research, 2-2-3 Minatojima-minamimachi, Chuo-ku, Kobe 650-0047, Japan}
\affiliation{RIKEN Cluster for Pioneering Research, 2-2-3 Minatojima-minamimachi, Chuo-ku, Kobe 650-0047, Japan}
\affiliation{Universal Biology Institute, The University of Tokyo, Bunkyo-ku, Tokyo 113-0033, Japan}

\date{\today}

\begin{abstract}
Motivated by the observations of intracellular phase separations and the wetting of cell membranes by protein droplets, we study the nonequilibrium surface wetting by Monte Carlo simulations of a lattice gas model involving particle creation.
We find that, even when complete wetting should occur in equilibrium, the fast creation of particles can hinder the surface wetting for a long time due to the bulk droplet formation.
Performing molecular dynamics simulations, we show that this situation also holds in colloidal particle systems when the disorder density is sufficiently high.
The results suggest an intracellular control mechanism of surface wetting by changing the speed of component synthesis.
\end{abstract}

\maketitle

\textit{Introduction.}
Liquid-liquid phase separation has recently drawn attention in the field of cell biology~\cite{Hyman2014, Shin2017, Banani2017}.
The physics and biological functions of phase separation have been elucidated for proteins and RNAs, including nucleoli~\cite{Brangwynne2011, Weber2015, Falahati2016}, nuclear bodies~\cite{Berry2015, Pak2016}, and disordered nuage proteins~\cite{Nott2015}, to name a few.
Particularly interesting is how the positioning of phase-separated droplets inside cells may be physically controlled by the wetting properties of the components.
Wetting properties have been found to be important in germline development of \textit{C. elegans}~\cite{Brangwynne2009}, regulation of the autophagy~\cite{Agudo2021}, and initiation of the endocytosis~\cite{Day2021}.
Moreover, it has been shown that the Par proteins in cultured cells of \textit{Drosophila} can first form droplets in the cytoplasm before wetting the cell membrane, which can be a key process in setting the polarity in asymmetric cell division~\cite{Oon2019, Kono2019, Liu2020, Wu2020}.

Surface wetting has been theoretically~\cite{deGennes1985} and experimentally~\cite{Tanaka2001, Bonn2009} studied based on free energy arguments, for instance, to understand the conditions of complete, partial, and non-wetting in equilibrium~\cite{Bonn2001}.
These theories are not directly applicable to the situation inside cells, since the intracellular environment is typically out of equilibrium due to the chemical reactions~\cite{Wurtz2018}.
In terms of the dynamics, the motion of the droplets is restricted in a size-dependent manner, as the cell is likely packed with organelles and cytoskeleton meshes~\cite{Arrio-Dupont2000}, and crowded with macromolecules~\cite{Weiss2004}.
Therefore, it is interesting to consider what strategies cells can be taking to localize liquid droplets in the cytoplasm or on the membranes controlling simple chemical kinetics such as protein creation and degradation.

In this Letter, we investigate how the phase diagram of surface wetting is affected when the components of the phase-separated droplets are created over time.
Using a simple lattice model with particle creation, we perform numerical simulations and obtain a generic formula for the wetting condition.
We find that, even under the situations where complete wetting should occur in equilibrium, frequent particle creation can prevent the surface wetting for a long time due to the initial nucleation event that takes place in the bulk.
Moreover, performing molecular dynamics (MD) simulations of colloidal particles with impurities mimicking the disorder effects in the cytoplasm, we show that the wetting condition obtained from the lattice model is applicable when the impurity density is large enough.
Based on these results, we find the condition window that the cells may be utilizing to achieve surface wetting.

\textit{Model and equilibrium phase diagram.}
As a minimal model of the intracellular phase separation, we consider interacting particles in a two-dimensional square lattice with lattice constant 1.
A particle at $(x, y)$ ($1 \leq x, y \leq L$) can hop to a neighboring empty site at a rate $D \min \{ 1, \mathrm{e}^{-\Delta E} \}$, where $\Delta E$ is the energy increase by hopping (in units of $k_\mathrm{B} T$).
We assume a nearest-neighbor interaction $-J \ (< 0)$, which can be largely negative to induce equilibrium phase separation, and $\Delta E = -J \Delta n$, where $\Delta n$ is the change in the number of nearest-neighbor bonds by hopping.
The effect of the membrane is represented by the one-dimensional boundaries at $y=0$ and $L+1$.
The affinity of the surface is parameterized by $-h$, which is the interaction between the surface and neighboring particles.
To reduce the geometrical effects of corners, we adopt the periodic boundary condition in the $x$-direction.

Motivated by the intracellular protein synthesis, we further assume that particles are generated in a randomly chosen empty site at a rate $\lambda$ $(<D)$, starting from the initial state with no particles.
In addition, we model the saturation of the protein concentration by stopping the particle creation when the particle density reaches a saturation value $\rho_\mathrm{sat}$ and set that time as $t = \tau_\mathrm{sat}$.
Thus, the time evolution of the mean density $\rho (t)$ is represented as $\rho (t) = 1 - \mathrm{e}^{-\lambda t} = 1 - (1 - \rho_\mathrm{sat})^{t / \tau_\mathrm{sat}}$ (for $t < \tau_\mathrm{sat}$) and $\rho (t) = \rho_\mathrm{sat}$ (for $t > \tau_\mathrm{sat}$).
The total time $t_\mathrm{tot} \ (> \tau_\mathrm{sat})$, which includes the waiting time after the saturation, is taken as the order of the cell-cycle period, representing the typical time scale that the intracellular environment changes.
We introduce the typical time of diffusion $\tau_\mathrm{diff} := L^2 / D$ and the typical creation time of a single particle $\tau_\mathrm{cre} := \tau_\mathrm{sat} / \rho_\mathrm{sat} L^2 \propto 1/\lambda$.
Then, the effective model parameters are $L$, $J$, $h$, $\rho_\mathrm{sat}$, $t_\mathrm{tot} / \tau_\mathrm{diff}$, and $\tau_\mathrm{cre} / \tau_\mathrm{diff}$.

To investigate the time evolution of the lattice gas model, we perform Monte Carlo simulations in the following way.
First, we randomly choose one of the $L^2$ sites and decide whether to hop in one of the four directions or create a particle with a probability of 1/5 each.
Then, the hopping or particle creation is performed when the target site is empty, with probability $D \min \{ 1, \mathrm{e}^{- \Delta E} \} \Delta t$ or $\lambda \Delta t$, respectively, where $\Delta t = D^{-1}$.
We count this single step as a time increment of $\Delta t / 5 L^2$, and repeat the procedure until the final time $t = t_\mathrm{tot}$.
To obtain the equilibrium phase diagram, we instead take a random configuration with density $\rho$ as the initial state and set $\lambda = 0$.

We note that this lattice model demonstrates diffusion-limited dynamics, meaning that the droplets can grow or shrink via Ostwald ripening but the motion of the liquid droplets is negligible compared to the diffusion of single particles. 
This type of model has been used in explaining phase separation kinetics observed in cells where the motion of the droplets tends to be very slow~\cite{Weiss2004, Lee2013, Shin2017b, Wurtz2018, Wurtz2018_2, Dine2018}, possibly due to the high density of cytoskeletons, organelles, and other macromolecules.

\begin{figure}[t]
\includegraphics[scale=0.5]{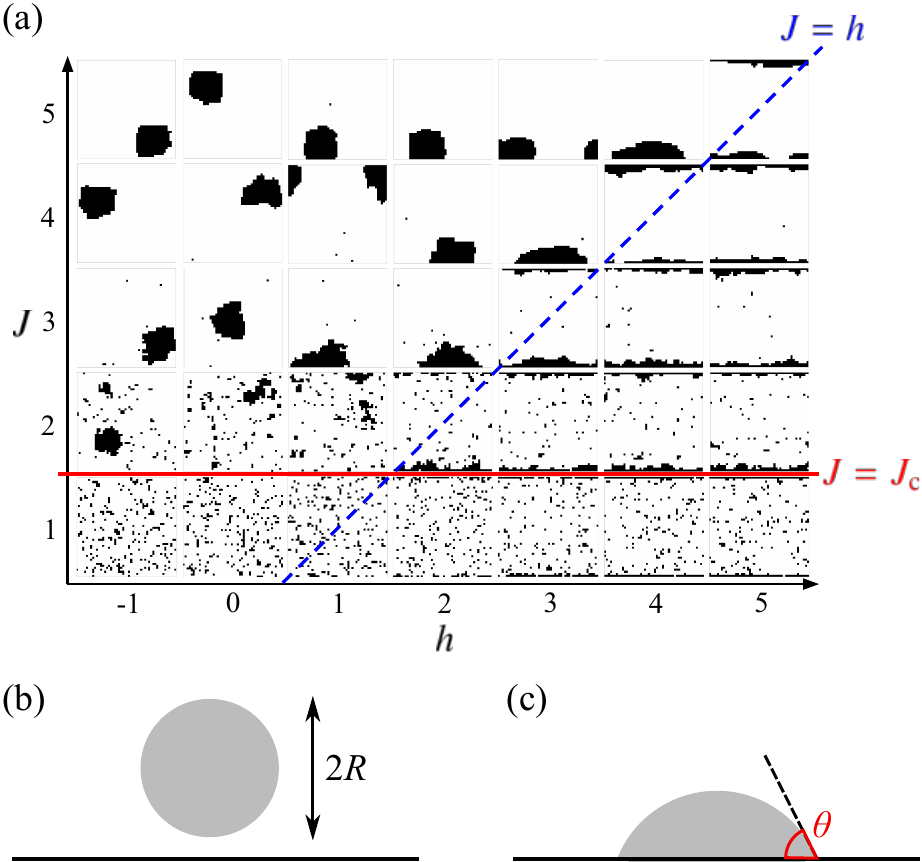}
\caption{
(a) Phase diagram of equilibrium state with representative snapshots of configurations.
The red solid line is the onset of phase separation, and the blue dashed line is the theoretical boundary between partial wetting and complete wetting.
We used $L = 50$, $\rho = 0.1$, and $t_\mathrm{tot} / \tau_\mathrm{diff} = 10000$ in the Monte Carlo simulation.
Schematic figures of (b)~a non-wetting circular droplet with radius $R$ and (c)~a partially wetting droplet with a wetting angle $\theta$ are also shown.}
\label{Fig:EqPhDiag}
\end{figure}

We first show in Fig.~\ref{Fig:EqPhDiag}(a) the equilibrium phase diagram of a system with size $L=50$ and density $\rho = 0.1$ with representative configurations after a long enough waiting without particle creation.
We can see that the phase-separated droplets are formed when $J$ is larger than a certain value $J_\mathrm{c}$ ($\sim 2$), which corresponds to the coexistence line.
In addition, the droplet is localized on the surfaces, or surfaces are wet, for sufficiently large $h$; especially for $h \gtrsim J$, the wetting angle is almost zero.

Assuming large $J$, we can derive the wetting conditions observed in Fig.~\ref{Fig:EqPhDiag}(a) by the following argument.
The surface energy of a circular droplet with radius $R$ [Fig.~\ref{Fig:EqPhDiag}(b)] is estimated as $\pi R  J$, while that of a droplet wetting the surface with an angle $\theta$ [Fig.~\ref{Fig:EqPhDiag}(c)] is $r(\theta) (\theta + \sin \theta) J - 2 r(\theta) h \sin \theta$, where $r (\theta)$ is the radius of curvature of the droplet.
Equality of the volume between the non-wetting and partially wetting droplets leads to $\pi R^2 = (\theta - \sin \theta \cos \theta) {r(\theta)}^2$, from which we can obtain the $\theta$ dependence of $r (\theta)$.
Minimizing the energy difference between the wet and ``dry'' conditions with respect to $\theta$, we can obtain the energetically favored states depending on $J$ and $h$: non-wetting for $h < 0$, partial wetting with $\cos \theta = (2 h - J) / J$ for $0 < h < J$, and complete wetting with $\theta = 0$ for $h > J$.

\textit{Kinetics-dependent wetting fraction.}
Next, we consider the situation with particle creation, with the wettable surface condition $h = J$, where the equilibrium state shows the complete wetting [see Fig.~\ref{Fig:EqPhDiag}(a)].
Since larger affinity can only lower the probability of particles detaching from the surface without changing the energetically favored configuration (complete wetting), the following results will not depend on the value of $h$ as long as $h \geq J$.

\begin{figure*}[t]
\includegraphics[scale=0.36]{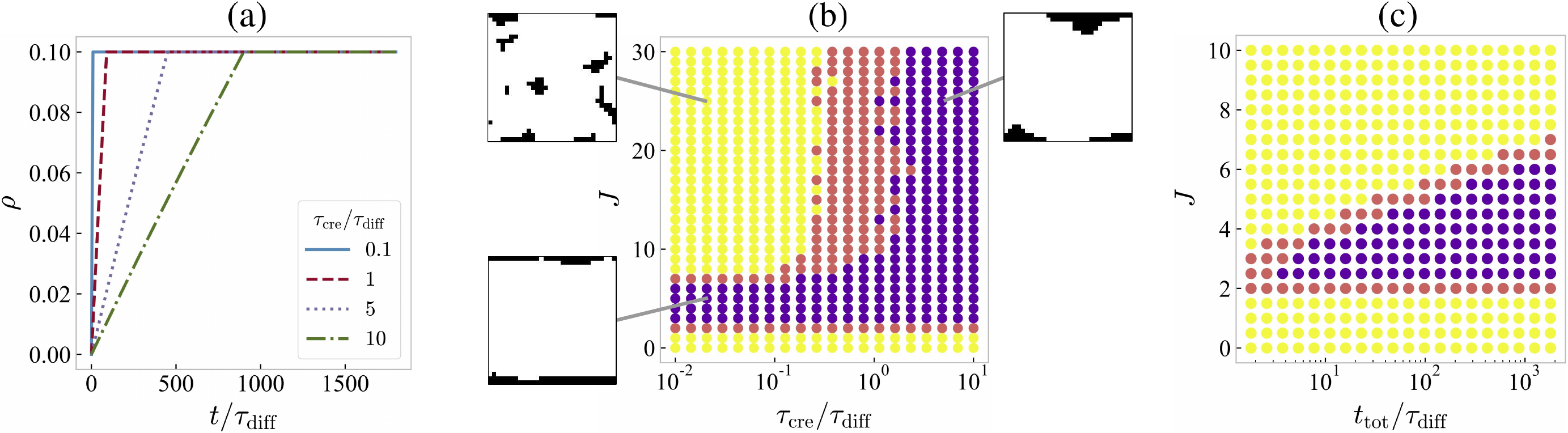}
\caption{
(a) Time evolution ($0 \leq t \leq t_\mathrm{tot} = 1800 \tau_\mathrm{diff}$) of the mean density $\rho$ for $\tau_\mathrm{cre} / \tau_\mathrm{diff} = 0.1$ (blue solid), 1 (red dashed), 5 (purple dotted), and 10 (green dash-dotted) with $\rho_\mathrm{sat} = 0.1$.
(b) The heatmap of the wetting fraction $\phi_\mathrm{w}$ at $t = t_\mathrm{tot} \ (= 1800 \tau_\mathrm{diff})$ (yellow for $\phi_\mathrm{w} \leq 0.7$, orange for $0.7 < \phi_\mathrm{w} \leq 0.9$, and purple for $\phi_\mathrm{w} > 0.9$) as a function of the creation time $\tau_\mathrm{cre}$ and the interaction strength $J$, with typical final configurations.
The value of $\phi_\mathrm{w}$ at each point is statistically averaged over 30 independent numerical simulations.
We used $L = 30$ and $\rho_\mathrm{sat} = 0.1$.
(c) The heatmap similar to (b) at $\tau_\mathrm{cre} / \tau_\mathrm{diff} = 0.01$ as a function of  the total time $t_\mathrm{tot}$ and the interaction strength $J$.
The value of $\phi_\mathrm{w}$ at each point is averaged over 100 independent numerical simulations.
The other parameters are the same as (b).
}
\label{Fig:WetFrac}
\end{figure*}

As an indicator of surface wetting, we use the wetting fraction $\phi_\mathrm{w}$, i.e., the fraction of particles that are in contact with the surface directly or indirectly through other particles.
Setting a long simulation time compared with the free-particle diffusion time ($t_\mathrm{tot} \gg \tau_\mathrm{diff}$), we fix the system size, the saturation density, and the total time as $L = 30$, $\rho_\mathrm{sat} = 0.1$, and $t_\mathrm{tot} = 1800 \tau_\mathrm{diff}$, respectively, while changing the interaction strength $J$ and the particle creation time $\tau_\mathrm{cre}$.
The time evolution of the mean density $\rho (t)$ for several values of $\tau_\mathrm{cre} / \tau_\mathrm{diff}$ is shown in Fig.~\ref{Fig:WetFrac}(a).
Note that the final state at $t = t_\mathrm{tot}$ is not necessarily at equilibrium.

The heatmap in Fig.~\ref{Fig:WetFrac}(b) shows the $J$ and $\tau_\mathrm{cre}$ dependence of $\phi_\mathrm{w}$ averaged over 30 independent samples.
For $J \lesssim 2$, $\phi_\mathrm{w}$ is low since phase separation does not occur in this region just as in the equilibrium state [see $J < J_\mathrm{c}$ in Fig.~\ref{Fig:EqPhDiag}(a)]; conversely, based on the equilibrium diagram, the complete wetting (with purple colors) is expected as long as $J \gtrsim 2$ is satisfied.
However, we find significant $J$ and $\tau_\mathrm{cre}$-dependence of $\phi_\mathrm{w}$ for $J \gtrsim 7$ (from yellow to purple via orange with decreasing $J$ or increasing $\tau_\mathrm{cre}$).
Since $t_\mathrm{tot}$ is now fixed, this result indicates that the approach to equilibrium becomes slower due to the rapid particle creation and the strong interactions.

To understand the observed behavior of $\phi_\mathrm{w}$ in Fig.~\ref{Fig:WetFrac}(b), we consider the condition where the surface wetting is achieved during the set time $t_\mathrm{tot}$.
As a sufficient condition for surface wetting, we first note that if a particle reaches the surface before another particle is created, the droplet will grow on the surface.
Since the diffusion time of a particle from bulk to the surface is $O(\tau_\mathrm{diff})$ and the particle creation time is $\tau_\mathrm{cre}$, this scenario will be achieved if
\begin{equation}
\tau_\mathrm{cre} > C \tau_\mathrm{diff},
\label{Eq:tau0}
\end{equation}
where $C = O(1)$.
This condition is consistent with the seemingly vertical boundary between the high-$\phi_\mathrm{w}$ region (purple) and the middle-$\phi_\mathrm{w}$ region (orange) in Fig.~\ref{Fig:WetFrac}(b).

To consider the effect of particle interactions, we next focus on the parameter region with $\tau_\mathrm{cre} \ll \tau_\mathrm{diff}$, where the droplets will grow in bulk.
Within the mean-field picture, a particle constituting a drop will freely hop $C_0 L^2 (1 - \rho_\mathrm{sat})$ times while colliding with other particles $C_0 L^2 \rho_\mathrm{sat}$ times on average until it reaches the surface [$C_0 = O(1)$ is a constant].
On the other hand, the time consumed during the collision will be $D^{-1} \mathrm{e}^{C_1 J}$ since the detaching rate of adjacent particles is $D \mathrm{e}^{-C_1 J}$ according to the dependence of hopping probability on energy increase [$C_1 = O(1)$ represents the mean coordination number].
Since the particles bound together can only diffuse at a negligible speed in this model, the effective diffusion time of a particle from bulk to the surface will be $\tilde{\tau}_\mathrm{diff} \sim D^{-1} C_0 L^2 (1 - \rho_\mathrm{sat}) + D^{-1} \mathrm{e}^{C_1 J} C_0 L^2 \rho_\mathrm{sat}$, or
\begin{equation}
\tilde{\tau}_\mathrm{diff} \sim C_0 \tau_\mathrm{diff} \left( 1 - \rho_\mathrm{sat} + \rho_\mathrm{sat} \mathrm{e}^{C_1 J} \right).
\label{Eq:tautil}
\end{equation}
If the effective diffusion time $\tilde{\tau}_\mathrm{diff}$ is shorter than the total time $t_\mathrm{tot}$, the particles will finally accumulate on the surface and the surface wetting will be achieved.
Since $t_\mathrm{tot} \gg \tau_\mathrm{diff}$, we can rewrite this condition for surface wetting as
\begin{equation}
J < {C_1}^{-1} \ln \left( \frac{t_\mathrm{tot}}{C_0 \rho_\mathrm{sat} \tau_\mathrm{diff}} \right).
\label{Eq:J0}
\end{equation}
This indicates that for droplets or aggregates formed by sufficiently strong interactions, the time it takes for the wetting can be exponentially long.

The wetting condition Eq.~\eqref{Eq:J0} can be checked by simulating the case with rapid particle creation ($\tau_\mathrm{cre} / \tau_\mathrm{diff} = 0.01$).
We plot in Fig.~\ref{Fig:WetFrac}(c) the $t_\mathrm{tot}$ and $J$ dependence of $\phi_\mathrm{w}$.
In the region satisfying $t_\mathrm{tot} / \tau_\mathrm{diff} \gg 1$, we can see that the upper boundary between the high-$\phi_\mathrm{w}$ (purple) and low-$\phi_\mathrm{w}$ (yellow) regions follows $J \propto \ln t_\mathrm{tot}$.

\textit{Effect of protein degradation.}
In a real cell, the saturation of a specific protein concentration may be caused by the balance between the production and degradation, rather than a monotonic increase.
Here we will consider how the annihilation of particles (by rate $\sigma$) will change the phase diagram of surface wetting.
Models of phase separation under particle creation and annihilation~\cite{Glotzer1994, Glotzer1994PRE, Glotzer1995, Lefever1995, Carati1997} have been recently discussed in the context of intracellular droplet formations~\cite{Wurtz2018, Wurtz2018_2, Lee2018, Berry2018}.
The time evolution of the mean density $\rho (t)$ in this case is given by $\rho (t) = (1 + \sigma / \lambda)^{-1} [1 - \mathrm{e}^{- (\lambda + \sigma) t}]$.
Therefore, we re-define $\rho_\mathrm{sat}$ and $\tau_\mathrm{sat}$ so that $\rho (t) = \rho_\mathrm{sat} (1 - \mathrm{e}^{- t / \tau_\mathrm{sat}})$. The typical time scale of particle creation is then consistently given by $\tau_\mathrm{cre} := \tau_\mathrm{sat} / \rho_\mathrm{sat} L^2 \propto 1/\lambda$.

\begin{figure}[t]
\includegraphics[scale=0.28]{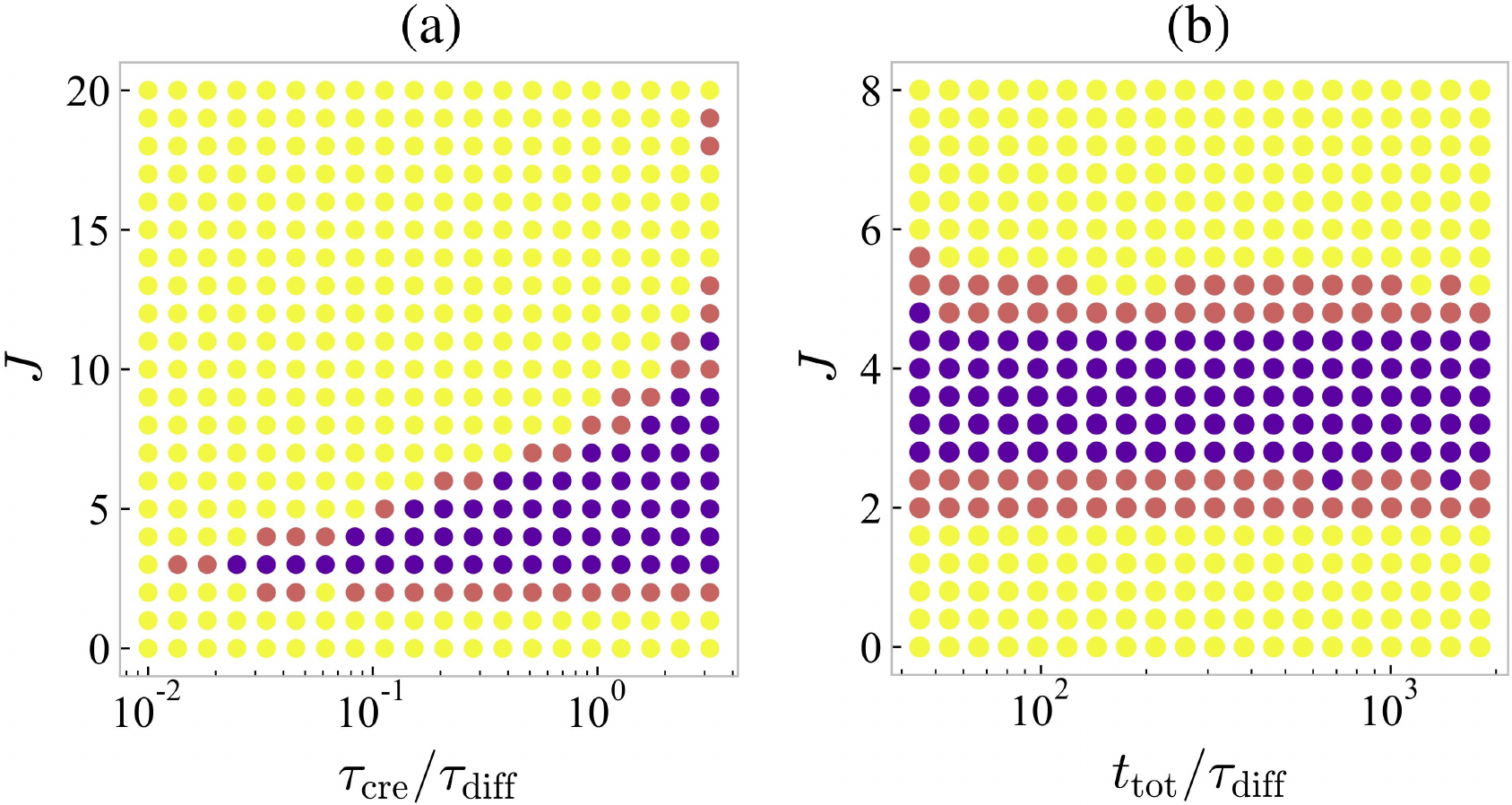}
\caption{
(a) A color plot of the wetting fraction $\phi_\mathrm{w}$ as a function of the creation/annihilation time $\tau_\mathrm{cre}$ and the interaction strength $J$ in the model including creation/annihilation processes [with colors used in the same way as Fig.~\ref{Fig:WetFrac}(b)].
We used $L = 30$, $\rho_\mathrm{sat} = 0.1$, and $t_\mathrm{tot} / \tau_\mathrm{diff} = 1800$.
(b) $\phi_\mathrm{w}$ as a function of the total time $t_\mathrm{tot}$ ($45 \leq t_\mathrm{tot} / \tau_\mathrm{diff} \leq 1800$) and the interaction strength $J$ for a fixed value of $\tau_\mathrm{cre} / \tau_\mathrm{diff} \ (= 0.1)$.
The other parameters are the same as in (a).
Each value of $\phi_\mathrm{w}$ in (a) and (b) is statistically averaged over 10 and 30 independent simulations, respectively.
}
\label{Fig:WFReac}
\end{figure}

In Fig.~\ref{Fig:WFReac}(a), the wetting fraction $\phi_\mathrm{w}$ at $t = t_\mathrm{tot}$ is plotted against $\tau_\mathrm{cre} / \tau_\mathrm{diff}$ and $J$.
Comparing Fig.~\ref{Fig:WFReac}(a) with Fig.~\ref{Fig:WetFrac}(b), we can see that the low-$\phi_\mathrm{w}$ region with yellow colors is extended when adding the effect of degradation. In particular, the upper boundary between the high-$\phi_\mathrm{w}$ (purple) and low-$\phi_\mathrm{w}$ (yellow) regions merges with the lower boundary.
This indicates that there is a maximum speed of particle creation that allows surface wetting to occur.

For $\tau_\mathrm{cre} / \tau_\mathrm{diff} \lesssim 1$ in Fig.~\ref{Fig:WFReac}(a), the steady state is achieved at $t = t_\mathrm{tot}$ due to fast creation/annihilation dynamics, as exemplified by the very weak $t_\mathrm{tot}$ dependence of $\phi_\mathrm{w}$ for the case of $\tau_\mathrm{cre} / \tau_\mathrm{diff} = 0.1$ [Fig.~\ref{Fig:WFReac}(b)].
If $\tau_\mathrm{cre}$ is longer than a typical nucleation time, which is given by~\cite{Sosso2016} $\tau_\mathrm{nucl} = C_2 \tau_\mathrm{diff} \mathrm{e}^{C_3 J}$ with some constants $C_2$ and $C_3$, the droplets will grow up and the surface wetting will occur.
Thus, the condition for surface wetting is given as
\begin{equation}
J < {C_3}^{-1} \ln \left( \frac{\tau_\mathrm{cre}}{C_2 \tau_\mathrm{diff}} \right).
\label{Eq:J1}
\end{equation}
With a small $C_2 \ (\simeq 0.1)$, Eq.~\eqref{Eq:J1} explains the upper boundary in Fig.~\ref{Fig:WFReac}(a), and the lowest $\tau_\mathrm{cre}$ to observe surface wetting can be obtained by $\tau_\mathrm{cre} \simeq C_2 \tau_\mathrm{diff} \mathrm{e}^{2 C_3}$.
Therefore, even in the steady state, the interaction-dependent surface wetting occurs as in the case of monotonic protein level increase, or the wetting is prohibited if the protein turnover is so fast.

\textit{Wetting condition for diffusive particles.}
\begin{figure}[t]
\includegraphics[scale=0.8]{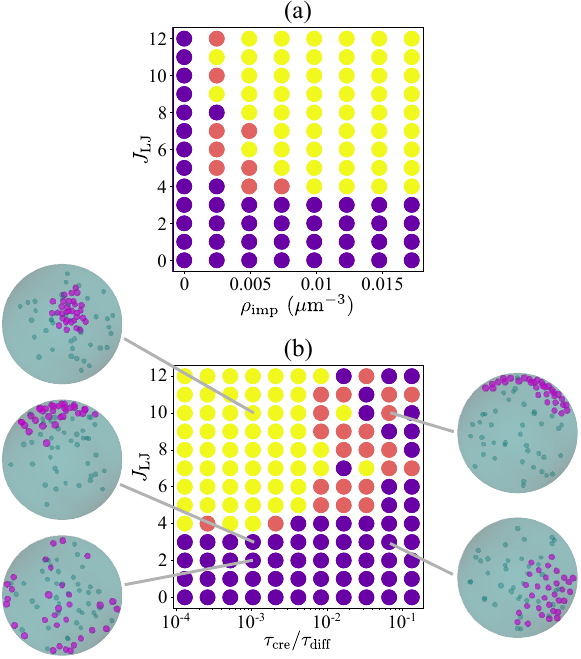}
\caption{
(a) The heatmap of the mean wetting fraction $\phi_\mathrm{w}$ at $t = t_\mathrm{tot} \ (=6.5 \tau_\mathrm{diff})$ (yellow for $\phi_\mathrm{w} \leq 0.85$, orange for $0.85 < \phi_\mathrm{w} \leq 0.95$, and purple for $\phi_\mathrm{w} > 0.95$) as a function of the density of disorder particles $\rho_\mathrm{imp}$ and the interaction strength of the Lennard-Jones potential $J_{\rm LJ}$.
(b) The same heatmap as a function of the creation time $\tau_\mathrm{cre}$ and the interaction strength of the Lennard-Jones potential $J_{\rm LJ}$ at $t = t_\mathrm{tot} \simeq 6.5 \tau_\mathrm{diff}$ for $\rho_\mathrm{imp} \simeq 0.0123 \mathrm{\mu m}^{-3}$, with typical final configurations.
The value of $\phi_\mathrm{w}$ at each point is statistically averaged over more than 50 independent numerical simulations.
}
\label{Fig:WetFracMD}
\end{figure}
We next checked whether similar situations arise in a lattice-free system: a three-dimensional model of diffusive particles, where not only single components but aggregates can also move freely~\cite{SM}.
We consider $N$ diffusive particles inside a cell-mimicking sphere, interacting with each other through the Lennard-Jones (LJ) potential.
To model the disorder effects in the cytoplasm caused for example by cytoskeleton meshes and macromolecular crowding, we further introduced disorder particles that are immobile in the sphere.
The disorder particles interact repulsively with the diffusive particles.

We first examined the wetting fraction $\phi_\mathrm{w}$ for quickly created particles ($\tau_\mathrm{cre} \to 0$) as a function of the impurity density $\rho_\mathrm{imp}$ and the LJ interaction strength $J_\mathrm{LJ}$ [Fig.~\ref{Fig:WetFracMD}(a)].
For small $\rho_\mathrm{imp}$, $\phi_\mathrm{w}$ is independent of $J_\mathrm{LJ}$, in contrast to the $J$-dependent $\phi_\mathrm{w}$ observed in the lattice simulation [$\tau_\mathrm{cre} / \tau_\mathrm{diff} \ll 1$ in Fig.~\ref{Fig:WetFrac}(b)].
This reflects that a droplet with the radius $R$ can move with the diffusivity $\sim D a / R$, distinct from the exponentially small diffusivity $\sim D \exp (-C_1 J)$ of adjacent particles in the lattice model.

On the other hand, for large $\rho_\mathrm{imp}$, $\phi_\mathrm{w}$ diminishes as $J_\mathrm{LJ}$ is increased in a similar way as observed in the lattice model.
We can interpret this behavior as follows.
For a droplet trapped by surrounding impurities to escape from the trap, interfacial particles constituting the droplet must diffuse away by overcoming the energy barrier of the LJ potential, which leads to the Arrhenius-type $J_\mathrm{LJ}$ dependence of the diffusivity, which is similar to the lattice case.
Consistently, we find that $\phi_\mathrm{w}$ increases as $\tau_\mathrm{cre}$ becomes larger [Fig.~\ref{Fig:WetFracMD}(b)].
Note that $\phi_\mathrm{w} \simeq 1$ even without droplets ($J_\mathrm{LJ} \lesssim 2$) because the particle number is so small that a single layer of particles is formed on the surface [see the bottom left configuration in Fig.~\ref{Fig:WetFracMD}(b)].
Thus, if the impurity density is large enough, the kinetics-dependent facilitation/suppression of surface wetting can occur also in colloidal systems as predicted in the lattice simulation.

\textit{Discussion and conclusion.}
To discuss the relevance of the effect of kinetics on the surface wetting in cells based on the obtained formula [Eqs.~\eqref{Eq:tau0} and \eqref{Eq:J0}], let us consider the typical value of the cell size as $L = 20 \, \mathrm{\mu m}$, the diffusion coefficient as $D = 10 \, \mathrm{\mu m^2 / s}$, and the cell-cycle period as $t_\mathrm{tot} = 24 \, \mathrm{hours}$ to represent the time scale of intracellular property change.
The diffusion time from the cytoplasm to the membrane is then estimated as $\tau_\mathrm{diff} = L^2 / D = 40 \, \mathrm{s}$.
The particle creation time, or the protein synthesis time, can be calculated as $\tau_\mathrm{cre} = \tau_\mathrm{sat} / N_\mathrm{sat}$ if the saturation time $\tau_\mathrm{sat}$ and the saturation number of proteins $N_\mathrm{sat}$ are given.
The ratio between the cell-cycle period and the diffusion time is $t_\mathrm{tot} / \tau_\mathrm{diff} = 2160$.

As a first example, let us consider a slow protein synthesis and take $\tau_\mathrm{sat} = 10 \, \mathrm{hours}$ and $N_\mathrm{sat} = 100$.
In this case, we obtain $\tau_\mathrm{cre} = 360 \, \mathrm{s}$ and $\tau_\mathrm{cre} / \tau_\mathrm{diff} = 9$, which, based on Eq.~\eqref{Eq:tau0}, satisfies the condition for surface wetting regardless of the interaction strength.
As another example, we consider a fast protein synthesis and take $\tau_\mathrm{sat} = 1 \, \mathrm{hour}$ and $N_\mathrm{sat} = 1000$.
Then, we obtain  $\tau_\mathrm{cre} = 3.6 \, \mathrm{s}$ and $\tau_\mathrm{cre} / \tau_\mathrm{diff} = 0.09$, which, based on Eq.~\eqref{Eq:J0}, is on the margin of surface wetting condition depending on the interaction strength.
As we have seen, particle annihilation by protein degradation can work to prevent surface wetting.
Therefore, we propose that wettability of cellular and nuclear membrane surfaces may be regulated using changes in protein-protein interactions by post-translational modifications and/or changes in the speed of protein synthesis by gene expression regulation.
The presented arguments should hold generically for three-dimensional cells since dimensionality does not affect the derivation of Eqs.~\eqref{Eq:tau0} and \eqref{Eq:J0}.

In this Letter, we have studied a simple model of phase separation in the presence of particle creation to consider the wetting conditions in the intracellular environment.
We have shown through the lattice model and theory that slow protein creation or moderate interaction strength is required for surface wetting to be achieved when the protein has a phase-separating property in the bulk.
We have confirmed by MD simulations that the same situation occurs also in diffusive particle systems in the presence of disorder, due to the effect of the trapping of the droplets.
Our results demonstrate how the liquid droplets and their locations in cells may not directly reflect the equilibrium phase, which is directly relevant in interpreting the images of dynamic cell membrane wetting.
It will be interesting to compare the phase diagram with experiments with controlled protein synthesis to elucidate the details of the intracellular environment.

\textit{Acknowledgments.}
We are grateful to F. Matsuzaki, K. Kono, T. Shibata, and I. Fujita for experimental suggestions and fruitful discussions, and H. Tanaka and H. Nakano for helpful comments.
This work was supported by JSPS KAKENHI grants number JP18H04760, JP18K13515, JP19H05275, JP19H05795, 19K16096, JP20K14435 and Research Grant from Human Frontier Science Program (Ref. Grant No. RGY0081/2019).

\newpage

\onecolumngrid

\renewcommand{\thepage}{S\arabic{page}}  
\renewcommand{\thesection}{S\arabic{section}}   
\renewcommand{\thefigure}{S\arabic{figure}}
\renewcommand{\theequation}{S\arabic{equation}}
\setcounter{page}{1}
\setcounter{section}{0}
\setcounter{figure}{0}
\setcounter{equation}{0}

\begin{center}
\textbf{ \large Supplemental Material for \\ Surface wetting by kinetic control of liquid-liquid phase separation}\\ [.1cm]
{Kyosuke Adachi and Kyogo Kawaguchi}\\ [.1cm]
{(Dated: \today)}\\
\end{center}

\section{Molecular dynamics simulation}
We employed molecular dynamics (MD) simulations using OpenMM~\cite{Eastman2017}.
The simulation consists of two types of particles; diffusive macroscopic particles that attract each other and can phase separate, and disorder particles that are static but can interact with the diffusive particles.
The diffusive particles undergo Brownian motion with diffusion constant $D=6.4  \, \mathrm{\mu m}^2/\mathrm{sec}$. The attractive interaction between the diffusive particles are given by the Lennard-Jones potential,
\begin{eqnarray}
U_\mathrm{df}(r)=4 J_\mathrm{LJ} k_\mathrm{B} T \left [ \left( \frac{\sigma}{r} \right )^{12}- \left( \frac{\sigma}{r} \right )^6 \right ],
\end{eqnarray}
where $r$ is the distance between the diffusive particles, $J_\mathrm{LJ}$ is the parameter setting the strength of the attractive interaction, $k_\mathrm{B}$ is the Boltzmann constant, and $T=310.15$ K.
Here, $\sigma=2 \, \mathrm{\mu m}$ was set large so that the effect of interest can be observed with smaller number of particles (i.e., low cost in numerics).
The interaction between the disorder particles and the diffusive particles is given by Weeks-Chandler-Andersen potential, which is the repulsive part of the Lennard-Jones potential:
\begin{eqnarray}
U_\mathrm{do}(r)= \begin{cases}
8 k_\mathrm{B} T \left [ \left( \frac{\sigma}{r} \right )^{12}- \left( \frac{\sigma}{r} \right )^6 \right ] &  (r \leq r_0) \\
8 k_\mathrm{B} T \left [ \left( \frac{\sigma}{r_0} \right )^{12}- \left( \frac{\sigma}{r_0} \right )^6 \right ] & (r>r_0)
\end{cases}
\end{eqnarray}
where $r$ is the distance between a diffusive particle and a disorder particle, and $r_0=2^{1/6} \sigma$ is the equilibrium length of the Lennard-Jones potential. 
We set the diffusion constant of the disorder particles as $10^{-5} D$ so that their motion is negligible.

The system is covered by a spherical shell with radius $L=9.9 \, \mathrm{\mu m}$, which represents the position of the cell membrane. The diffusive particles tend to stick to the shell due to the potential:
\begin{eqnarray}
U_\mathrm{s} (r_i)= \begin{cases}
k_\mathrm{B} T & (r_i<L-\sigma)\\
 k_\mathrm{B} T  (r_i-L)^2 & (L-\sigma \leq r_i < L)\\
5 k_\mathrm{B} T (r_i-L)^2 & (L \leq r_i ).
\end{cases}
\end{eqnarray}
where $r_i$ is the distance between the origin and the particle. The typical time for a diffusive particle to reach the shell is $\tau_\mathrm{diff} = L^2/D = 15.3 \, \mathrm{sec}$.

In the simulations, we placed $N_\mathrm{d}= \frac{4}{3} \pi L^3 \rho_\mathrm{imp} $ disorder particles at random positions inside the shell.
Then, we started generating the diffusive particles one by one at random positions inside the sphere with diameter 0.77$\,L$ with time interval $\tau_\mathrm{cre}$. The time increment of the simulation was $2.5 \times 10^{-5}\, \mathrm{sec}$. After $N$ particles were generated, we continued the simulation up to the set total time $t_{\rm tot} =100\, \mathrm{sec}$ (4 $\times 10^6$ steps). 

We iterated the simulation for each condition more than 50 times to obtain the phase diagrams (Fig.~4 in the main text). The mean wetting fraction was calculated at the final time frame ($t=t_{\rm tot}$) by counting the number of diffusive particles near the surface of the sphere (i.e., $L-1.5 \sigma \leq r_i$) and all the particles that were connected to those particles. Here, the particles were counted as connected when they were within the distance of $1.5 \sigma$.

%

\end{document}